\documentclass[12pt]{article}
\usepackage{epsf}

\usepackage{scicite}

\usepackage{times}

\def\ga{\lower.5ex\hbox{$\; \buildrel > \over \sim \;$}}
\def\la{\lower.5ex\hbox{$\; \buildrel < \over \sim \;$}}
\def\etal{{\it et al.}}

\def\mG{~\mu{\rm G}}
\def\nG{~{\rm nG}}
\def\kms{~{\rm km~s^{-1}}}
\def\Mpc{~h^{-1}{\rm Mpc}}

\newenvironment{sciabstract}{%
\begin{quote} \bf}
{\end{quote}}

\newcounter{lastnote}
\newenvironment{scilastnote}{%
\setcounter{lastnote}{\value{enumiv}}%
\addtocounter{lastnote}{+1}%
\begin{list}%
{\arabic{lastnote}.}
{\setlength{\leftmargin}{.22in}}
{\setlength{\labelsep}{.5em}}}
{\end{list}}

\title{Turbulence and Magnetic Fields
in the Large Scale Structure of the Universe}

\author{
Dongsu Ryu,$^{1\ast}$ Hyesung Kang,$^{2}$ Jungyeon Cho,$^{1}$
and Santabrata Das$^{3}$ \\
\\
\normalsize{$^{1}$Department of Astronomy and Space Science,
Chungnam National University,}\\
\normalsize{Daejeon 305-764, Korea}\\
\normalsize{$^{2}$Department of Earth Sciences,
Pusan National University, Pusan 609-735, Korea}\\
\normalsize{$^{3}$Astrophysical Research Center for the Structure and
Evolution of the Cosmos,}\\
\normalsize{Sejong University, Seoul 143-747, Korea}\\
\\
\normalsize{$^\ast$To whom correspondence should be addressed.
E-mail: ryu@canopus.cnu.ac.kr}
}

\date{}

\begin{document} 


\baselineskip24pt


\maketitle 


\begin{sciabstract}

The nature and origin of turbulence and magnetic fields in the
intergalactic space are important problems that are yet to be
understood.
We propose a scenario in which turbulent flow motions are
induced via the cascade of the vorticity generated at
cosmological shocks during the formation of the large scale structure. 
The turbulence in turn amplifies weak seed magnetic fields of any
origin.
Supercomputer simulations show that the turbulence is
subsonic inside clusters/groups of galaxies, whereas it is
transonic or mildly supersonic in filaments.
Based on a turbulence dynamo model, we then estimate that the
average magnetic field strength would be a few $\mG$ inside
clusters/groups, approximately $0.1 \mG$ around clusters/groups,
and approximately $10 \nG$ in filaments. 
Our model presents a physical mechanism that transfers the gravitation
energy to the turbulence and magnetic field energies in the large
scale structure of the universe. 

\end{sciabstract}

\clearpage



There is growing evidence that the intergalactic medium (IGM) is
permeated with magnetic fields and is in a state of turbulence,
similar to the interstellar medium within galaxies.
Magnetic fields in the intracluster medium (ICM) have been measured
using a variety of techniques, including observations of diffuse
synchrotron emission from radio halos, inverse-Compton scattered
cosmic background radiation in extreme ultraviolet and hard x-ray
radiation, and Faraday rotation measure (RM).
The inferred strength of the magnetic fields is on the order of
$1 \mG$ \cite{ct02,gf04,ckb01}.
In the IGM outside of clusters, an upper limit of $\sim 0.1\mG$ has
been placed on the magnetic field strength of filaments, based on the
observed limit of the RMs of background quasars \cite{rkb98,xkhd06}.

So far, signatures of turbulence have been observed only in the ICM.
The analysis of the gas pressure maps of the Coma cluster revealed that
pressure fluctuations are consistent with Kolmogoroff turbulence,
and turbulence is likely to be subsonic with 
$\varepsilon_{\rm turb} \ga 0.1 \varepsilon_{\rm th}$, 
where $\varepsilon_{\rm turb}$ and $\varepsilon_{\rm th}$ are 
the turbulence and thermal energy densities, respectively
\cite{sfmb04}.
These results agree with predictions of numerical simulations 
of large scale structure (LSS) formation \cite{kcor97,nvk07}.
Turbulence in the ICM also has been studied in RM maps of a few
clusters \cite{mgfg04,ve05}.

It has been suggested that cosmological shocks 
with Mach numbers up to $\sim 10^4$ and 
speeds up to a few thousand ${\rm km~s^{-1}}$
exist in the IGM \cite{rkhj03,psej06,krco07}. 
Such shocks result from the supersonic flow motions that are
induced by the hierarchical formation of LSS in the universe.
They are collisionless shocks, which form in a tenuous plasma via
collective electromagnetic interactions between particles and
electromagnetic fields \cite{quest88}.
The gravitational energy released during the structure formation
is transferred by these shocks to the IGM plasma in several
different forms: in addition to the gas entropy,
cosmic rays are produced via diffusive shock acceleration
\cite{bell78,bo78},
magnetic fields are generated via the Biermann battery mechanism
\cite{biermann50,kcor97} and Weibel instability \cite{weibel59,msk06},
and vorticity is generated at curved shocks \cite{binn74,dw00}.

In astrophysical plasmas in which charged particles are coupled to
magnetic fields, turbulent flow motions and magnetic fields are
closely related.
We suggest that the turbulence in the IGM is induced by the
cascade of the vorticity generated at cosmological shocks.
The turbulence then amplifies the intergalactic magnetic fields
(IGMFs) through the stretching of field lines, a process known as
the turbulence dynamo.
This scenario provides a theoretically motivated model 
for the evolution of the IGMFs in LSS, independent of the origin of
seed fields.

There are other sources that can also provide turbulence and magnetic
fields to the IGM. 
For instance, galactic winds can drag out the galactic magnetic fields
on the order of $1 \mG$ strength into the surrounding IGM \cite{va99}.
The magnetic fields in the lobes of the jets from galactic black holes
can also contaminate the IGM \cite{kdlc01}.
Mergers of smaller objects are expected to produce turbulent motions
in the ICM, which in turn amplify the existing magnetic fields
\cite{ssh06}.
Those processes, although possibly important, are not topics of
this study.


We first calculated the vorticity,
${\vec \omega} \equiv {\vec\nabla}\times{\vec v}$ (curl of flow velocity),
in the IGM, from a numerical simulation using particle-mesh/Eulerian
hydrodynamic code \cite{rokc93} for the formation of LSS in a cold dark
matter dominated universe with a cosmological constant [supporting
online material (SOM) text S1].
As shown in Fig. 1, numerous shocks exist in the LSS that are bounded by
accretion shocks \cite{rkhj03}.
The distribution of vorticity closely matches that of shocks,
suggesting that a substantial portion of the vorticity, if not all,
must have been generated at the shocks.

There is a clear trend that the vorticity is larger in hotter 
(Fig. 2) and denser (fig. S1) regions.
As shown in the top right panel of Fig. 2, at the present epoch,
$\omega_{\rm rms}t_{\rm age} \sim 10$ to $30$ ($\omega_{\rm rms}$,
the root mean square of the vorticity; $t_{\rm age}$, the present
age of the universe) in clusters/groups (temperature $T > 10^7$ K)
and filaments ($10^5 < T < 10^7$ K), whereas it is on the order of
unity in sheetlike structures ($10^4 < T < 10^5$ K) and even smaller
in voids ($T < 10^4$ K) (see SOM text S2 for the temperature
phases of the IGM).
It increases a little with time and asymptotes after redshift $z \la 1$.
Because the local eddy turnover time, $t_{\rm eddy}$, can be defined
with the vorticity as $t_{\rm eddy} = 1/\omega$, 
$\omega t_{\rm age}(z)$ represents the number of eddy
turnovers in the age of the universe at a given $z$.
Roughly, if $\omega t_{\rm age}$ is greater than a few, we expect
there has been enough time for the vorticity to cascade down to
smaller scales and for turbulence to develop in the IGM.
So it is likely that turbulence is well developed in clusters/groups
and filaments, but the flow is mostly non-turbulent in sheets and voids.

In our simulation the vorticity was generated either directly at
curved cosmological shocks or by the baroclinity of flows.
The baroclinity resulted from the entropy variation induced at shocks.
Therefore, the baroclinic vorticity generation also can be attributed
to the presence of cosmological shocks. 
Our estimates of vorticity generation by the two processes
(SOM text S3) are shown with open symbols in the top right panel
of Fig. 2.
They agree reasonably well with the vorticity present in the
simulation, although the estimates are intended to be rough. 
The plot indicates that the contributions from the two processes
are comparable.

To estimate the energy associated with turbulence, 
the curl component of flow motions, ${\vec v}_{\rm curl}$, which
satisfies the relation
${\vec\nabla}\times{\vec v}_{\rm curl} \equiv {\vec\nabla}\times{\vec v}$,
is extracted from the velocity field (SOM text S4).
As vorticity cascades to develop into turbulence, the energy
$(1/2)\rho v_{\rm curl}^2$ ($\rho$, gas density) is transferred to
turbulent motions, so we regard it as the turbulence energy,
$\varepsilon_{\rm turb}$.
As shown in Fig. 3, 
$\varepsilon_{\rm turb} < \varepsilon_{\rm th}$ in clusters/groups.
In particular, the mass-averaged value is $\left<\varepsilon_{\rm turb}
/\varepsilon_{\rm th}\right>_{\rm mass} = 0.1$ to $0.3$ for $T > 10^7$K,
which is in good agreement with the observationally inferred value in
cluster cores \cite{sfmb04}.
The turbulence Mach number
$M_{\rm turb} \equiv v_{\rm turb}/c_{\rm s} = \sqrt{1.8} \
(\varepsilon_{\rm turb} / \varepsilon_{\rm th})^{1/2}$,
where $c_{\rm s}$ is the sound speed.
Therefore, overall turbulence is subsonic in clusters/groups,
whereas it is transonic or mildly supersonic in filaments.


The general consensus regarding the origin of the IGMFs is that no
mechanism can produce strong coherent magnetic fields in the IGM
before the formation of LSS and galaxies \cite{kz07}.
However, it is reasonable to assume that weak seed fields were
created in the early universe (SOM text S5).
The seed fields can be amplified by the intergalactic turbulence
discussed above.
In principle, if we were to perform magnetohydrodynamic (MHD)
simulations of structure formation, the amplification of the IGMFs
could be followed.
In practice, however, the currently available computational resources
do not allow a numerical resolution high enough to reproduce the full 
development of MHD turbulence in LSS \cite{kcor97}.

In order to follow the growth of the IGMFs by the dynamo action of
turbulence, we turned to a separate simulation in a controlled box.
Starting with a very weak regular field, a three-dimensional
incompressible simulation of driven MHD turbulence was performed
(SOM text S6).
In the simulation, the evolution of magnetic fields goes through
three stages:
(i) the initial exponential growth stage, when the back-reaction of
magnetic fields is negligible;
(ii) the linear growth stage, when the back-reaction starts to operate;
and (iii) the final saturation stage \cite{cv00}.
Adopting the simulation result, we model the growth and saturation
of magnetic energy as
$$
\phi(t/t_{\rm eddy}) = {\varepsilon_B \over \varepsilon_{\rm turb}}
= \cases{
0.04 \times \exp\left[(t/t_{\rm eddy}-4)/0.36\right]
& for\ $t/t_{\rm eddy}<4$ \cr
(0.36/41) \times (t/t_{\rm eddy}-4) + 0.04
& for\ $4<t/t_{\rm eddy}<45$ \cr
0.4
& for\ $t/t_{\rm eddy}>45$ \cr}
\eqno{(1)}
$$
(fig. S2).
Assuming that the fraction of turbulence energy governed by Eq. 1,
$\phi$, is converted into the magnetic energy, we estimate the
strength of the IGMFs as
$B=\left[8\pi\varepsilon_{\rm turb} \cdot \phi(\omega t_{\rm age})
\right]^{1/2}$.
Here the values of $\omega$ and $\varepsilon_{\rm turb}$ are
calculated locally from the structure formation simulation.

The resulting IGMFs follows the cosmic web of matter distribution
as shown in Fig. 4 (and in fig. S3). 
On average the IGMFs are stronger in hotter (Fig. 2) and denser 
(fig. S1) regions in our model. 
The strength of the IGMFs is $B \ga 1 \mG$ inside
clusters/groups (the mass-averaged value for $T > 10^7$ K),
$\sim 0.1 \mG$ around clusters/groups (the volume-averaged value for
$T > 10^7$ K), and $\sim 10 \nG$ in filaments at present (bottom right
panel of Fig. 2) (see SOM text S7 for the numerical convergence
of the estimation).
These values agree with the observed field strengths discussed earlier.
They also agree with the previous study \cite{kcor97}, in which the
magnetic field strength in clusters was estimated to be a few $\mu$G,
based on a kinetic theory.
The IGMFs should be much weaker in sheetlike structures and voids.
But as noted above, turbulence is not fully developed in such low
density regions, so our model is not adequate to 
predict the field strength there.
For each temperature phase, the IGMFs are stronger in the past,
because the gas density is higher.
However, the IGMFs averaged over the entire computational volume are
weaker in the past because the fraction of strong field regions
is smaller.

While being amplified, magnetic fields become coherent
through the inverse cascade \cite{cv00}.
The coherence scale of magnetic fields in fully developed turbulence
is expected to be several times smaller than the driving scale,
that is, the scale of dominant eddies (SOM text S8).
In the IGM outside of clusters, the curvature radius of typical
cosmological shocks is approximately a couple of Mpc
\cite{rkhj03} (fig. S4), which should represent a characteristic
scale of dominant eddies.
The coherence length of the IGMFs there is then expected to be
several 100 kpc.
On the other hand, the scale height of the ICM is several 100 kpc.
If it corresponds to the scale of the dominant eddies, the coherence
length in the ICM is expected to be $\sim 100$ kpc or so.

Our model can predict the RMs owing to the IGMFs, which may be tested
in future observations with Low Frequency Array and Square Kilometer
Array \cite{gbf04}.
Also, our model IGMFs can be employed in the study of the propagation
of ultra-high-energy cosmic rays, which is crucial to search for
astrophysical accelerators of such high energy particles \cite{dkrc08}.

\clearpage


\begin{figure}
\vspace{0cm}
\centerline{\epsfxsize=16cm\epsfbox{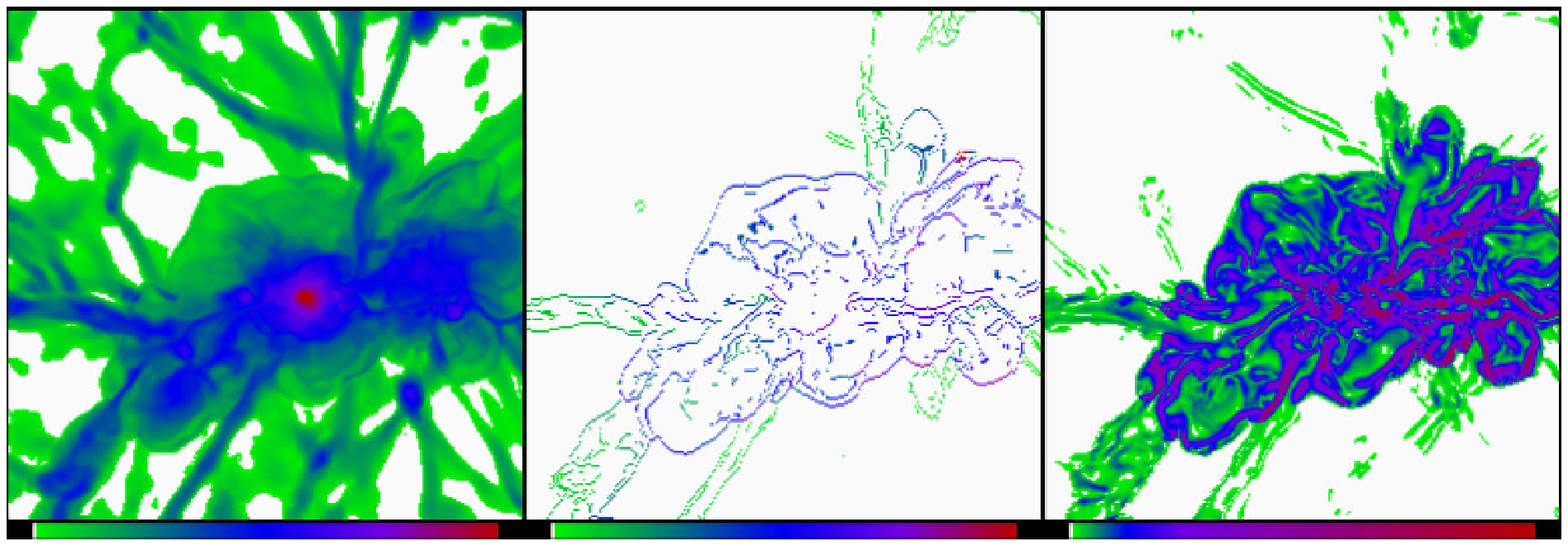}}
\vspace{0.5cm}
{Fig. 1. Two-dimensional images showing gas density $\rho$ in a
logarithmic scale (left), locations of shocks with color-coded
shock speed $v_{\rm shock}$ (middle), and magnitude of vorticity
$\omega t_{\rm age}$ (right), around a cluster complex of
$(25 \Mpc)^2$ area at present ($z=0$).
Here, $h$ is the Hubble constant in units of $100 \kms {\rm Mpc}^{-1}$.
The complex includes a cluster of x-ray emission-weighted temperature
$T_x \approx 3.3$ keV.
Color codes for each panel are (left) $\rho/\left< \rho \right>$ from
$10^{-1}$ (green) to $10^4$ (red); (middle) $v_{\rm shock}$ from 15
(green) to 1,800 $\kms$ (red); and (right) $\omega t_{\rm age}$ from
0.5 (green) to 100 (red).}
\end{figure}

\clearpage

\begin{figure}
\vspace{0cm}
\centerline{\epsfxsize=16cm\epsfbox{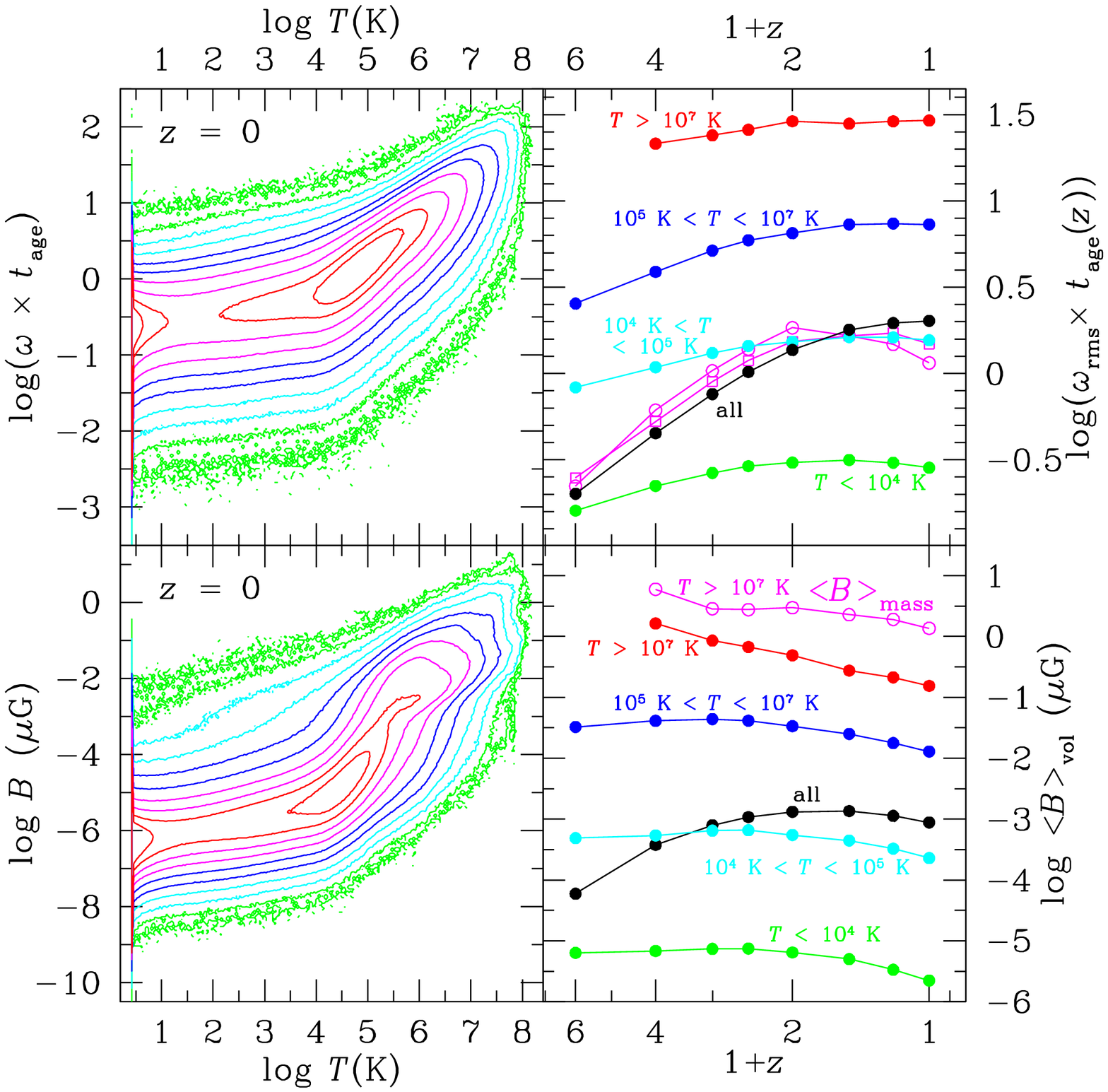}}
\vspace{0cm}
{Fig. 2. (Left) Volume fraction with given temperature and
vorticity magnitude (top left) and temperature and magnetic field
strength (bottom left) at present.
(Right) Time evolution of the root mean square of the vorticity
(top right) and the volume-averaged magnetic field strength
(bottom right) for four temperature phases of the IGM and for all
the gas as a function of redshift $z$.
Magenta symbols in the top right panel are our estimates of the
vorticity generated directly at curved shocks (open circles) and
by the baroclinity of flows (open squares).
Magenta open circles in the bottom right panel show the mass-averaged
magnetic field strength for $T > 10^7$ K.}
\end{figure}

\clearpage

\begin{figure}
\vspace{-4cm}
\centerline{\epsfxsize=16cm\epsfbox{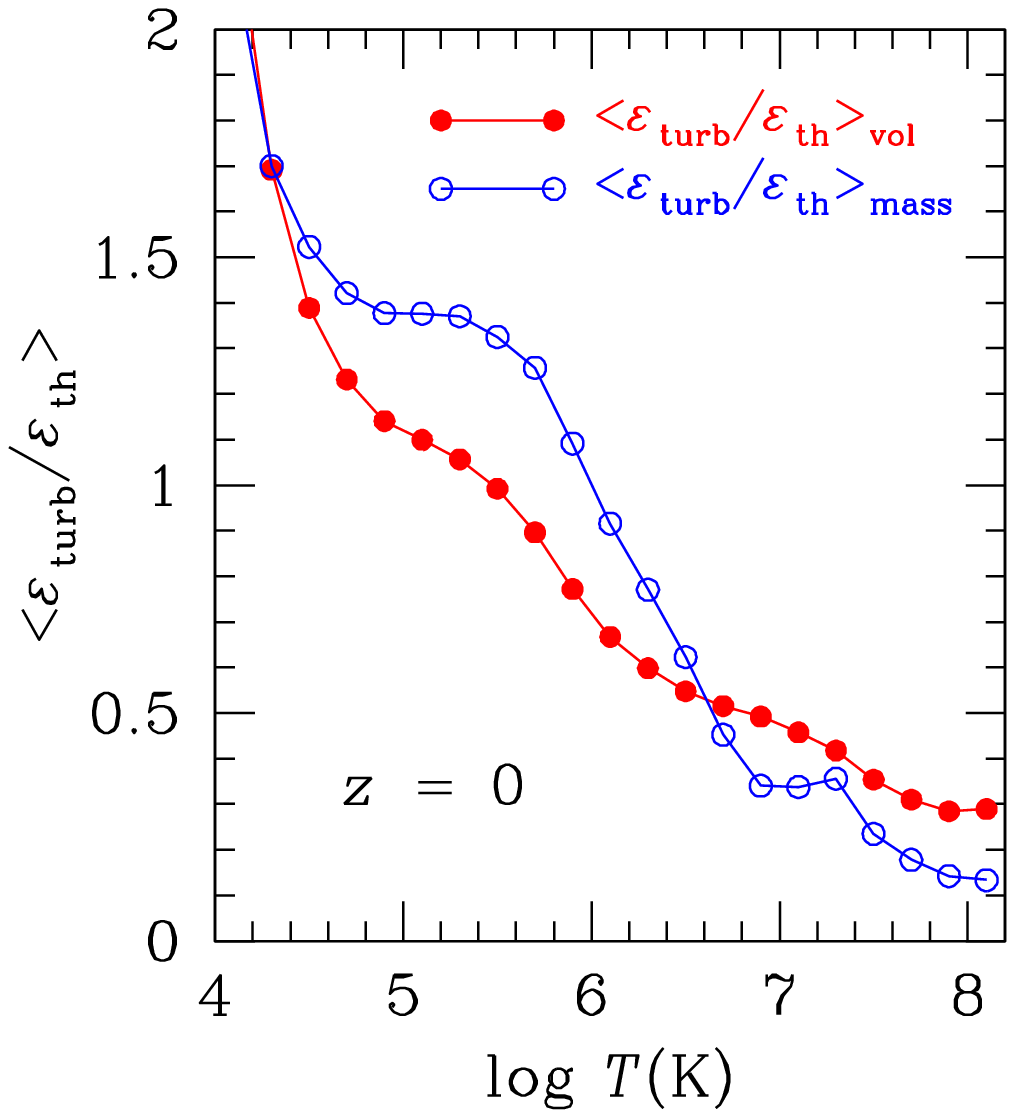}}
\vspace{-2.5cm}
{Fig. 3. Ratio of turbulence to thermal energies as a function of
temperature at present.
The values shown are volume-averaged and mass-averaged over
temperature bins.}
\end{figure}

\clearpage

\begin{figure}
\vspace{0cm}
\centerline{\epsfxsize=11cm\epsfbox{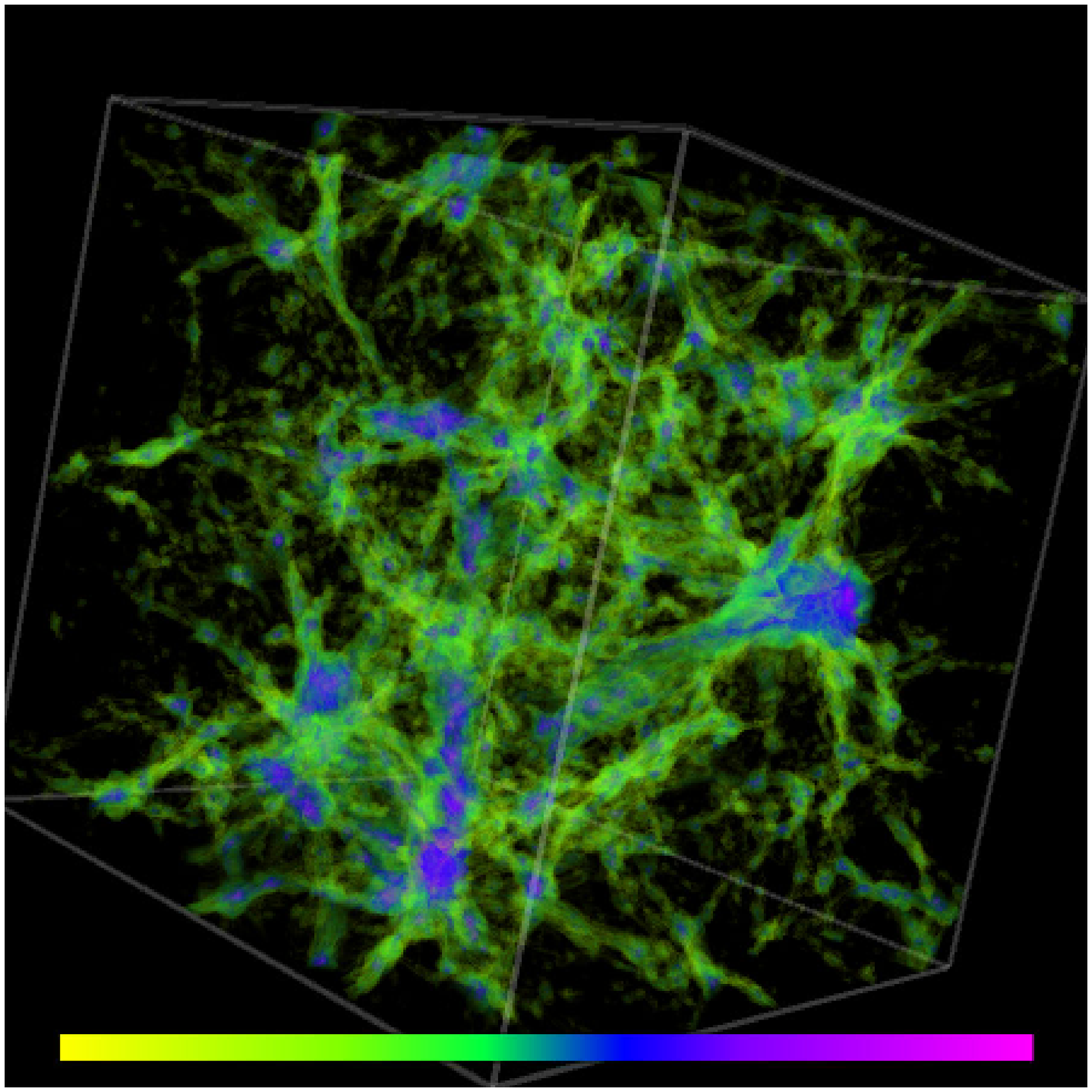}}
\vspace{0.5cm}
{Fig. 4. Volume-rendering image showing the logarithmically scaled
magnetic field strength at $z=0$ in the whole computational box of
$(100 \Mpc)^3$ volume.
Color codes the magnetic field strength from $0.1 \nG$ (yellow) to
$10 \mG$ (magenta).
The colors were chosen so that clusters and groups show as
magenta and blue and filaments as green.}
\end{figure}

\clearpage


\begin{scilastnote}
\item This work was supported by KICOS through grant
K20702020016-07E0200-01610 provided by MOST.
The work of D.R. was also supported by KOSEF through the grant of
basic research program R01-2007-000-20196-0.
The work of H.K. and S.D. was also supported by KOSEF through
ARCSEC.
The work of J.C. was also supported by KRF through grant
KRF-2006-331-C00136 funded by the Korean Government.
\end{scilastnote}

\clearpage

\centerline{\large \bf Supporting Online Material}

\vspace{3mm}

\noindent{\large \bf Materials and Methods}

\vspace{3mm}

\noindent{\bf S1. Simulation of large scale structure formation}
\hfill\break
The simulation used to estimate the vorticity in the intergalactic
medium (IGM) is for a $\Lambda$CDM universe with the following values
of cosmological parameters:
$\Omega_{BM}=0.043$, $\Omega_{DM}=0.227$, $\Omega_{\Lambda}=0.73$,
$h \equiv H_0$/(100 km/s/Mpc) = 0.7, and $\sigma_8 = 0.8$.
It was performed using a PM/Eulerian hydrodynamic code for the formation
of large scale structure (LSS) ({\it S1}), in a cubic region of comoving
volume $(100 \Mpc)^3$ with $1024^3$ grid zones for gas and gravity and
$512^3$ particles for dark matter, allowing a spatial resolution of
$\Delta l = 97.7$ kpc.
The simulation is adiabatic in the sense that it does not include
radiative cooling, galaxy/star formation, feedbacks from
galaxies/stars, and reionization of the IGM.
Although these processes play important roles in determining the
physical conditions mainly in cluster cores and voids, it was
shown that they do not affect significantly the global properties
and dissipations of cosmological shocks ({\it S2, S3}).

\vspace{3mm}

\noindent{\bf S2. Phases of the IGM}
\hfill\break
The intergalactic gas is heated mainly by cosmological shocks
({\it S4, S5}).
It was suggested that the IGM can be classified roughly into four
phases according to gas temperature: 
the hot gas with $T > 10^7$K mostly inside and around clusters/groups,
the warm-hot intergalactic medium (WHIM) with $T = 10^5 - 10^7$K
found mostly in filaments, the low temperature WHIM with
$T = 10^4 - 10^5$K distributed mostly as sheetlike structures,
and the diffuse gas with $T < 10^4$K residing mostly in voids
({\it S5, S6}).
In particular, the intracluster medium (ICM) refers to the hottest
phase of the IGM inside clusters that is observable in x-ray.

\vspace{3mm}

\noindent{\bf S3. Vorticity generation}
\hfill\break
Vorticity can be generated directly at curved shocks and by the
baroclinity of flows.
For uniform upstream flow, the vorticity produced behind curved
shock surface is
$$
{\vec \omega}_{\rm cs} = \frac{(\rho_2 - \rho_1)^2}{\rho_1 \rho_2}
K {\vec U}_1 \times {\hat n},
\eqno{(S1)}
$$
where $\rho_1$ and $\rho_2$ are the upstream and downstream gas
densities, respectively, ${\vec U}_1$ is the upstream flow velocity
in the shock rest frame, $K$ is the curvature tensor of the shock
surface, and ${\hat n}$ is the unit vector normal to the surface
({\it S7}).
If isopycnic surfaces (constant density surfaces) do not coincide
with isobaric surfaces, vorticity is generated with the rate given
by
$$
{\dot{\vec \omega}}_{\rm bc} = \frac{1}{\rho^2}
{\vec\nabla}\rho\times{\vec\nabla}p.
\eqno{(S2)}
$$

Rough estimates of the vorticity generations by the two processes
in the simulation of LSS formation are made as follows.
Considering the generation is a random walk process,
$$
({\dot\omega}\Delta t)_{\rm rms}
\left(\frac{t_{\rm age}(z)}{\Delta t}\right)^{1/2}
\eqno{(S3)}
$$
is calculated.
Here $\Delta t$ is the duration of coherent generation, and so
$t_{\rm age}(z)/\Delta t$ is the number of random walks during
the age of the universe at redshift $z$.
For the direct generation at cosmological shocks,
$$
{\dot\omega}_{\rm cs} \sim \frac{(\rho_2 - \rho_1)^2}{\rho_1 \rho_2}
\frac{U_{n1} U_{n2}}{R}\frac{g}{\Delta l}
\eqno{(S4)}
$$
is used.
Here $R$ is the curvature radius of the shocks, $U_{n1}$ and $U_{n2}$
are the flow speeds normal to the shocks in upstream and downstream,
$\Delta l$ is the grid size, and 
$g$ takes account of all other factors such
as geometric ones.
For the baroclinic vorticity generation,
${\dot\omega}_{\rm bc}$ from Eq. ({\it S2}) is used.
In the estimations, the distribution and property of shocks and flows
from the simulation are used.
Also $\Delta t$ is taken to be $10^8$ yrs, which is a typical
dynamical time, e.g., the sound crossing time in cluster cores.
The comoving value of $\left<R\right> \sim 1 \Mpc /(1+z)$, and
$\left<g\right> \sim 1$ are taken.
Note that $1 \Mpc$ is the typical scale of nonlinear structures
at present.
The factor $1/(1+z)$ is included because the average curvature radius
of cosmological shocks is smaller in the past even in the comoving
frame ({\it S8}).
In Fig. 2 the quantities normalized with $1/t_{\rm age}(z)$ are shown.

\vspace{3mm}

\noindent{\bf S4. Decomposition of flow velocity}
\hfill\break
The flow velocity can be decomposed into
$$
{\vec v} = {\vec v}_{\rm div} + {\vec v}_{\rm curl} +{\vec v}_{\rm unif},
\eqno{(S5)}
$$
where the divergence and curl components are defined as
${\vec\nabla}\cdot{\vec v}_{\rm div} \equiv {\vec\nabla}\cdot{\vec v}$ and
${\vec\nabla}\times{\vec v}_{\rm curl} \equiv {\vec\nabla}\times{\vec v}$,
respectively.
That is, ${\vec v}_{\rm div}$ is associated to compressional motions,
while ${\vec v}_{\rm curl}$ to incompressible shear motions.
Here ${\vec v}_{\rm unif}$ is the component uniform across the
computational box, whose magnitude is much smaller than the other
two components. 
The decomposition is calculated exactly in Fourier space.
We note with the above decomposition, locally
$$
{\vec v}_{\rm div}\cdot{\vec v}_{\rm curl} \ne 0
~~~~~~~~{\rm so}~~~~~~~~
\frac{1}{2} v^2 \ne
\frac{1}{2} \left(v_{\rm div}^2+ v_{\rm curl}^2 \right).
\eqno{(S6)}
$$
However, globally
$$
\int_{\rm box} {\vec v}_{\rm div}\cdot{\vec v}_{\rm curl} d^3{\vec x} = 0
~~~~~~~~{\rm so}~~~~~~~~
\int_{\rm box} \frac{1}{2} v^2 d^3{\vec x} = \int_{\rm box}
\frac{1}{2} \left(v_{\rm div}^2+ v_{\rm curl}^2 \right) d^3{\vec x}.
\eqno{(S7)}
$$

\vspace{3mm}

\noindent{\bf S5. Seed magnetic fields}
\hfill\break
A number of mechanisms that can create seeds field in the early
universe have been suggested.
Besides various inflationary and string theory mechanisms, the most
popular astrophysical mechanism is the Biermann battery ({\it S9}).
It was discussed in the context of cosmological shocks ({\it S10})
and ionization fronts ({\it S11}), and shown to build up
weak fields of strength up to $10^{-20}$ G by $z \sim$ a few.

At cosmological shocks, in addition, Weibel instability can
operate and produce magnetic fields up to the level of
$\varepsilon_B \sim 10^{-3} \varepsilon_{\rm sh}$ ({\it S12}),
and streaming cosmic rays accelerated by the shocks can amplify
weak upstream magnetic fields up to the level of 
$\varepsilon_B \sim (1/2)(U_1/c) \varepsilon_{\rm CRs}$
via non-resonant growing mode ({\it S13}).
Here $\varepsilon_{\rm sh}$ is the energy density of upstream flow,
and $\varepsilon_B$ and $\varepsilon_{\rm CRs}$ are the
energy densities of downstream magnetic fields and CRs.
With $U_{1}/c \sim 10^{-3}$ for cosmological shocks
({\it S2}, {\it S3}), these processes can potentially produce stronger
seed fields, although the coherence length of the resulting fields
is expected to very small and microscopic.

\vspace{3mm}

\noindent{\bf S6. Simulation of MHD turbulence}
\hfill\break
First we note there are at least two distinct types of
magnetohydrodynamic (MHD) turbulence:
one with strong regular fields and the other with weak/zero regular
fields.
The former has been described successfully with the nonlinear
interactions of Alfv\'en waves ({\it S14}). 
The latter is mainly hydrodynamic in large scales, 
but is modified by dynamically
important magnetic fields in small scales ({\it S15}, {\it S16}).
It is the latter that is relevant to this work.

Incompressible, driven turbulence was simulated initially with a
very weak regular field.
A pseudo-spectral code ({\it S15}) was used, employing hyper-viscosity
and hyper-resistivity with the Prandtl number of unity.
The advantages of this numerical approach include virtually zero
intrinsic numerical viscosity and resistivity, and the maximized
inertial range. 
The simulation was performed with a numerical resolution of $256^3$
collocation points.
The turbulence was driven at the scale of $L_{\rm driving} \sim
(1/2)L_{\rm box}$, where $L_{\rm box}$ is the computational box size.
The driving strength was set so that the total turbulence energy
becomes $\varepsilon_{\rm turb} \equiv \varepsilon_{\rm kin} +
\varepsilon_B \sim 1$ at saturation.
Initially $\varepsilon_B=10^{-6}$, but the evolution is not sensitive
to the initial field strength as long as it is sufficiently weak
({\it S16}).
The left panel of Fig. S2 shows the time evolution of kinetic and 
magnetic energies.
Here, the eddy turnover time is defined with the vorticity at driving
scale at saturation, $t_{\rm eddy} \equiv 1/\omega_{\rm driving}$.

We made the incompressible simulation, because compressible simulations
need much higher resolution to achieve the same growth rate and
magnetic field strength at saturation.
In addition, due to large numerical dissipation, controlling
viscosity and resistivity is not trivial in compressible simulations.
Nevertheless, in order to show the general behavior of turbulence
dynamo seen in our incompressible simulation  remains the same in
compressible regime, we performed compressible simulations of driven
MHD turbulence, using a code based on the third-order essentially
non-oscillatory (ENO) upwind scheme ({\it S17}).
This is one of the numerical schemes for MHD with least numerical
dissipation.
Different numerical resolutions of $8^3$ to $216^3$ grid zones
were used.
But otherwise, the initial conditions, the driving of turbulence,
and hyper-viscosity and hyper-resistivity with the Prandtl number
of unity were the same as in the incompressible simulation.
We only considered the case with the Mach number $M_s \sim 1$, which
is most relevant to this work.
We note that turbulence dynamo can be suppressed 
in highly supersonic flows ({\it S18}).

Fig. S5 compares the growth rate (left panel) and the magnetic
energy at saturation (right panel) from compressible simulations
with those from incompressible simulation.
Two points are clear from the figure: (1) the growth rate is
slower in compressible simulations, but the pattern follows that
in incompressible turbulence,
and (2) compressible simulation would need more than $1000^3$
grid zones to achieve the saturated magnetic field strength of
incompressible simulation.
It also demonstrates why it is impossible to reproduce the full 
development of MHD turbulence in the simulations of LSS formation
with the currently available computational resources.

\vspace{3mm}

\noindent{\bf S7. Convergence of our results}
\hfill\break
In order to test the numerical convergence of our estimation for
the strength of the intergalactic magnetic fields (IGMFs), we
repeated  the same analysis for the simulations of LSS formation
with lower numerical resolutions of $64^3$ to $512^3$ grid zones.
Except the resolution, they are the same simulations as that in
\S S1, in the sense that they have the same box size, the same
realization of initial conditions, and the same physics included.
Fig. S6 shows the volume-averaged magnetic field strength with
different resolutions.
The averaged strength around clusters/groups with $T > 10^7$K is
well converged at the resolution of $1024^3$.
On the other hand the averaged strength in filaments with
$10^5 < T < 10^7$K would be underestimated.
The converged value may be $\sim 2 - 3$ times larger than our
estimate with the resolution of $1024^3$.

\vspace{3mm}

\noindent{\bf S8. Coherence length}
\hfill\break
The consideration of the coherence length of turbulent magnetic
fields raises the issues of the scale of energy equipartition
and the magnetic field structure at the scale.
The right panel of Fig. S2 shows the power spectra for flow
velocity and magnetic fields at a time of saturation from the
simulation in \S S6.
The energy equipartition occurs at $\sim (1/2.5) L_{\rm driving}$,
which is close to the scale of the magnetic energy peak,
$\sim (1/3) L_{\rm driving}$.
The kinetic energy peak occurs at the the driving scale,
$L_{\rm driving}$.
Above the equipartition scale the flow structure is nearly isotropic,
while below it both the flow and magnetic field structures are
anisotropic and the eddies are stretched along the local magnetic
field lines ({\it S15}).
On the other hand, the magnetic field has most power at
$\sim (1/6) L_{\rm driving}$.
So we may argue that in the fully developed stage of the MHD
turbulence considered here, the coherence length of magnetic fields
would be several times smaller than the driving scale, the scale
of dominant eddies.

\vspace{3mm}

\noindent{\bf Additional Figures Used in Manuscript}

\vspace{3mm}

\noindent
Fig. S1 shows the correlation between the magnitude of vorticity
and the gas density (left panel) and the strength of magnetic fields
and the gas density (right panel) at present.

\vspace{3mm}

\noindent
Fig. S3 shows the spatial distribution of our model IGMFs around
a cluster complex and along a filament.
In the left image, in addition to the main cluster of
$T_x \approx 3.3$ keV, a group has $B \sim$ a few $\mu$G.
They are surrounded by a broad region of $B \sim 10\nG$. 
In the right image, groups with $B \sim \mG$ are distributed
along the filament with $B \sim 10\nG$. 

\vspace{3mm}

\noindent
Fig. S4 shows the power spectra for the flow velocity and its curl
and divergence components (\S S4) at present from the simulation of
LSS formation in \S S1.
At long wavelengths, the amplitude of perturbations are small, so
that the linear theory applies.
That is, $P_{\rm curl}(k) \rightarrow 0$  as $k \rightarrow 0$,
while $P_{\rm div}(k)$ follows the analytic theory expectation,
$P_{\rm div}(k) \sim k^{-1}$.
For wavelengths smaller than a few $\Mpc$, nonlinearities dominate,
and we see that $P_{\rm curl}(k) \ga P_{\rm div}(k)$.
$P_{\rm curl}(k)$ peaks at $\sim 5\Mpc$, and for values of $k$ somewhat
larger than the peak wavenumber, the spectrum follows a power law of
$k^{-5/3}$, the Kolmogorov spectrum.
$P_{\rm curl}(k)$ has most power at $\sim 2\Mpc$, that indicates the
typical scale of nonlinear structures in the simulation.

\clearpage

\begin{figure}
\vspace{-4cm}
\centerline{\epsfxsize=16cm\epsfbox{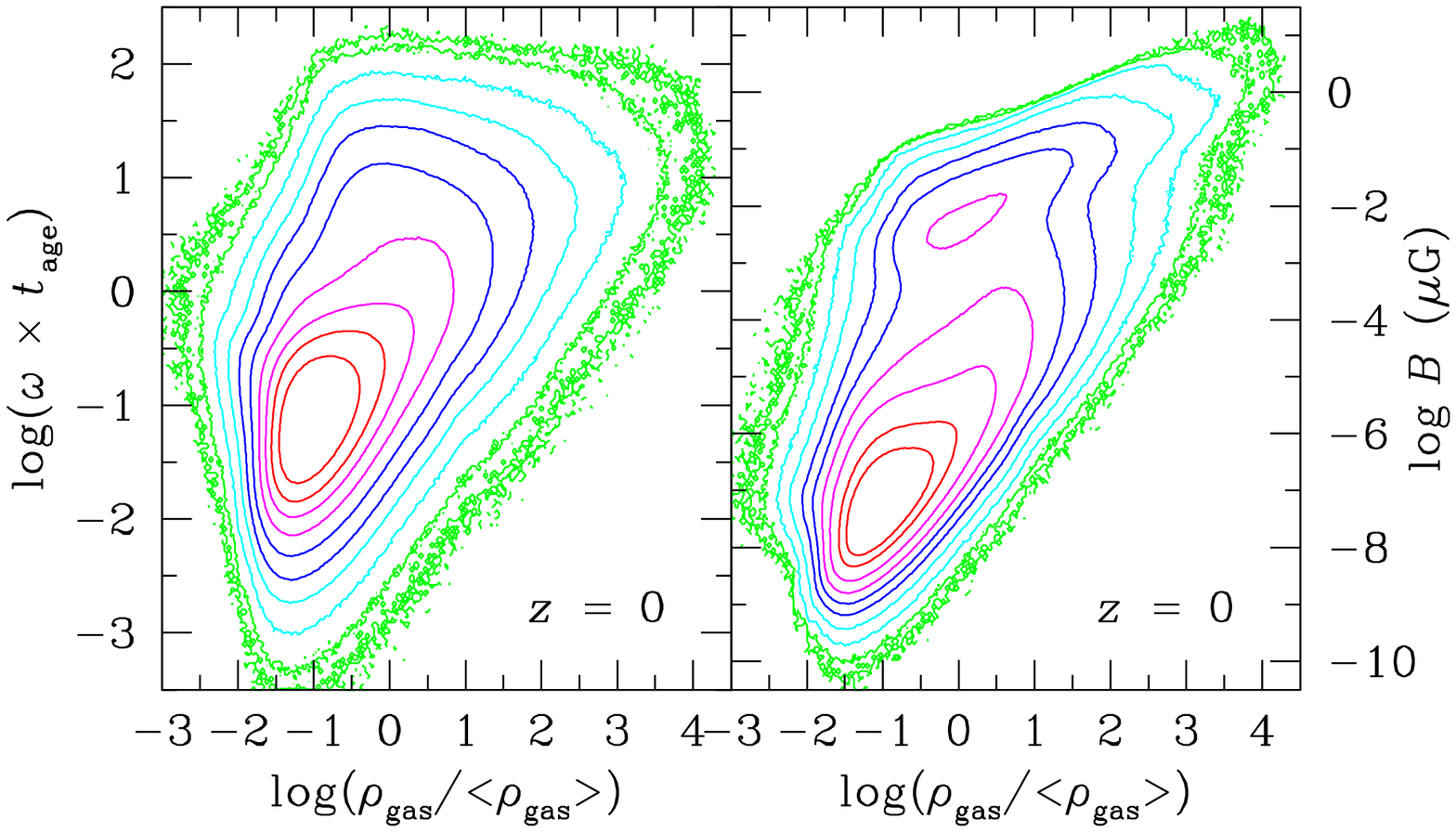}}
\vspace{-2.5cm}
{Fig. S1. Volume fraction with given gas density and vorticity
magnitude (left panel) and gas density and magnetic field strength
(right panel) at present.}
\end{figure}

\clearpage

\begin{figure}
\vspace{-4cm}
\centerline{\epsfxsize=16cm\epsfbox{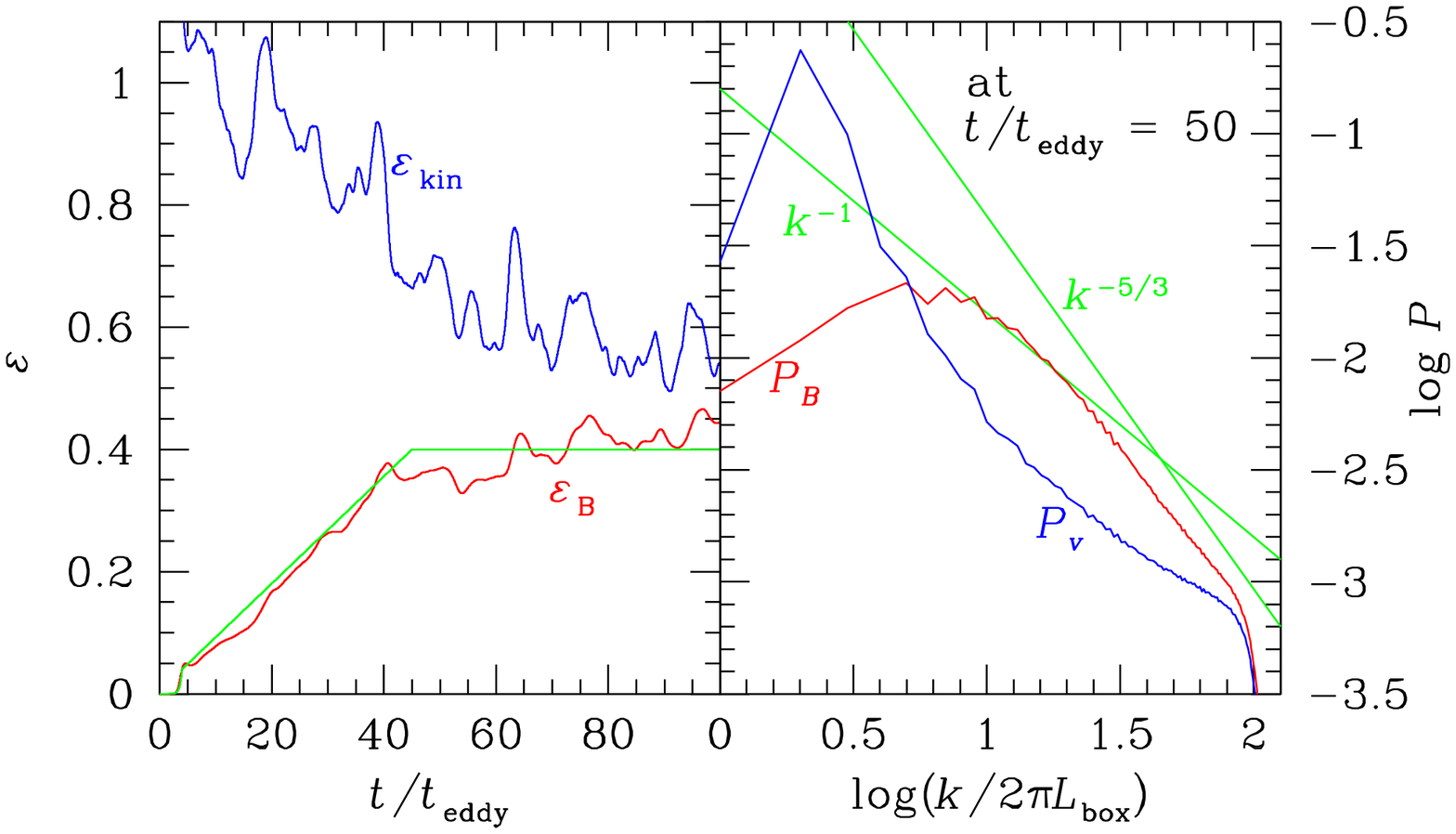}}
\vspace{-2.5cm}
{Fig. S2. Left panel: Time evolution of kinetic and magnetic
energies in a three-dimensional, incompressible simulation of driven
MHD turbulence with very weak initial magnetic field.
The green lines show our fitting for the growth and saturation of
magnetic energy.
Right panel: Power spectra for flow velocity, $P_v$, and magnetic
fields, $P_B$, at a time of saturation.
Two straight lines of slopes $-5/3$ and $-1$ are also drawn for
comparison.}
\end{figure}

\clearpage

\begin{figure}
\vspace{0cm}
\centerline{\epsfxsize=16cm\epsfbox{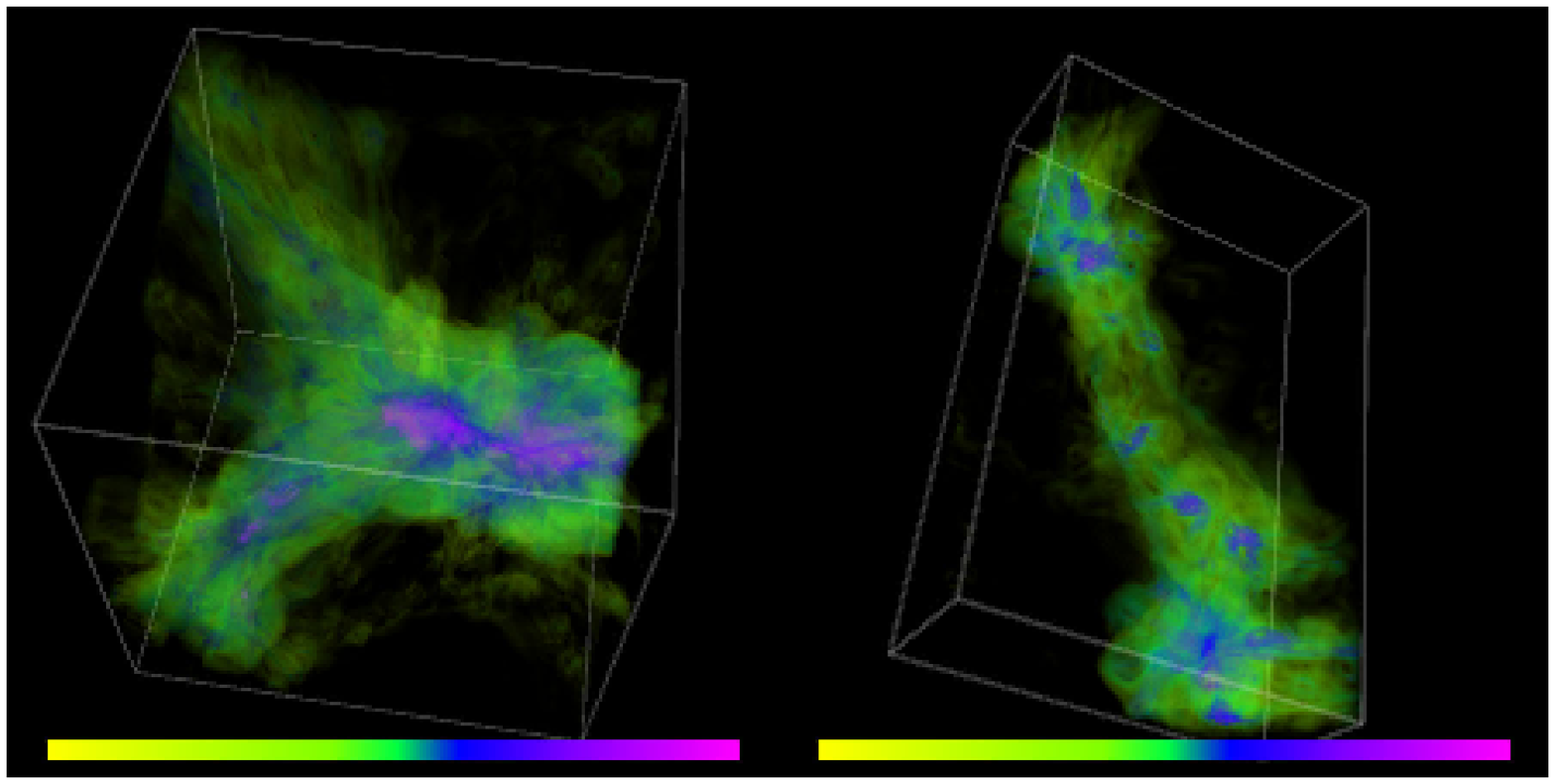}}
\vspace{0.5cm}
{Fig. S3. Volume rendering images showing logarithmically scaled
magnetic field strength at $z=0$ in a volume of $(25 \Mpc)^3$ around
the same cluster complex as shown in Fig. 1 (left panel) and in a
volume of $25\times15.6\times6.25\ (h^{-1}{\rm Mpc})^3$ which includes
a number of groups along a filament (right panel).
As in Fig. 4, color codes the magnetic field strength from $0.1 \nG$
(yellow) to $10 \mG$ (magenta).
Clusters and groups are shown with magenta and blue while filaments
with green.}
\end{figure}

\clearpage

\begin{figure}
\vspace{-4cm}
\centerline{\epsfxsize=16cm\epsfbox{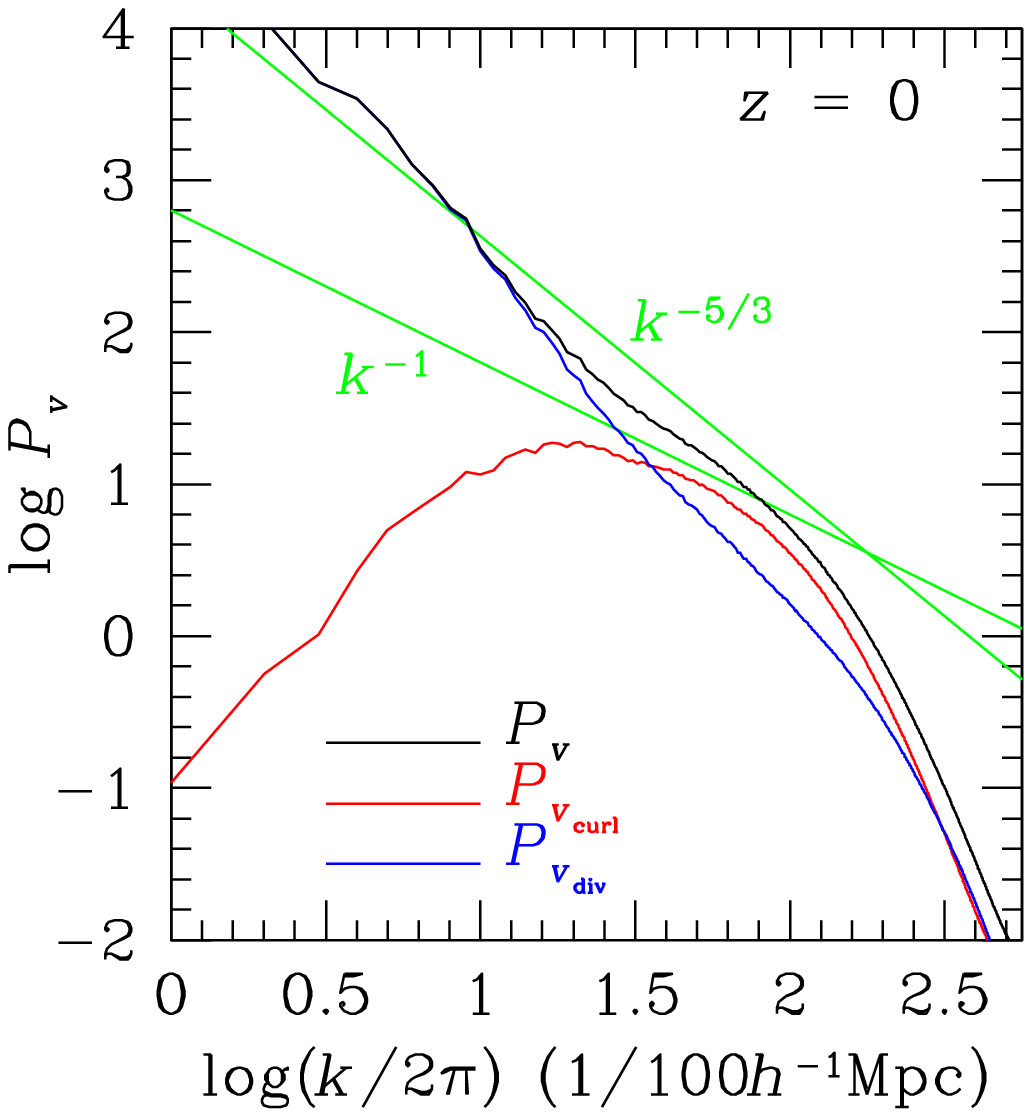}}
\vspace{-2.5cm}
{Fig. S4. Power spectra, $\int P_v dk = \left<(1/2) v^2 \right>$,
for flow velocity and its curl and divergence components at present
in the simulation of LSS formation.
Two straight lines of slopes $-5/3$ and $-1$ are also drawn for
comparison.}
\end{figure}

\clearpage

\begin{figure}
\vspace{-4cm}
\centerline{\epsfxsize=16cm\epsfbox{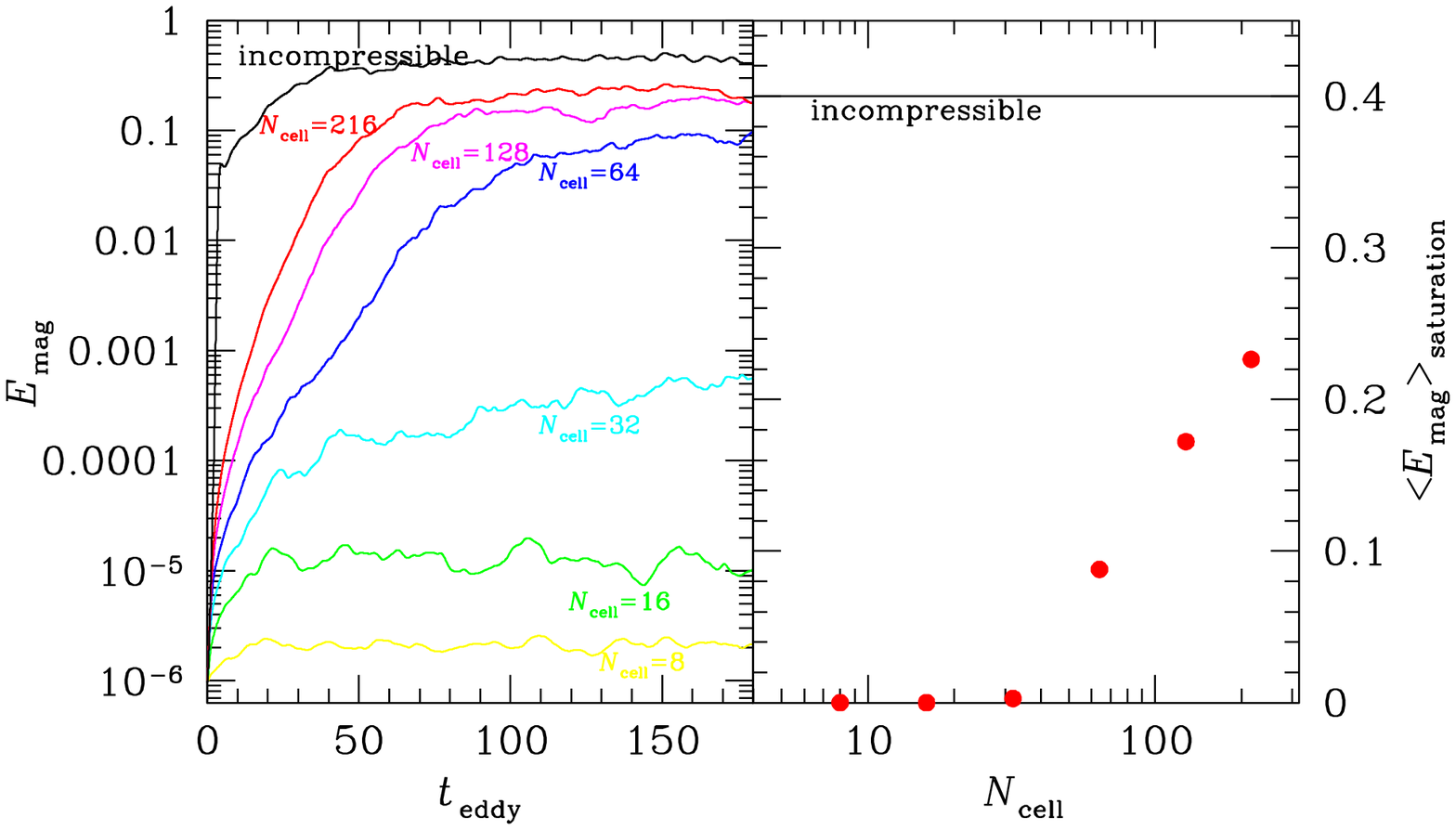}}
\vspace{-2.5cm}
{Fig. S5. Left panel: Time evolution of magnetic energy in
three-dimensional, compressible simulations of driven MHD
turbulence with very weak initial magnetic field.
For comparison, the evolution in incompressible simulation shown
in Fig. S3 is also plotted.
Right panel: Magnetic energy at saturation in compressible
simulations with different resolutions of $N_{\rm cell}^3$.
The value in incompressible simulation is marked with solid line.}
\end{figure}

\clearpage

\begin{figure}
\vspace{-4cm}
\centerline{\epsfxsize=16cm\epsfbox{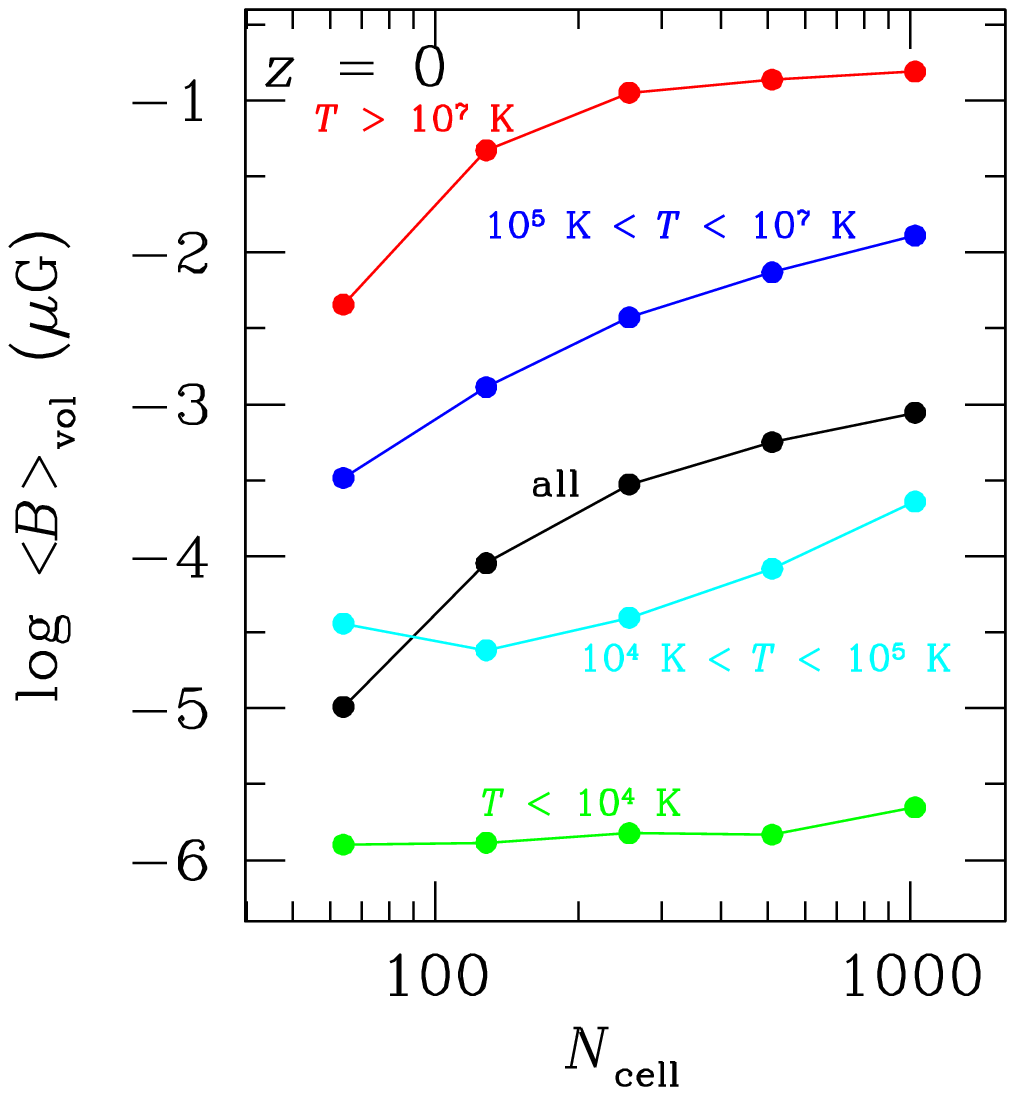}}
\vspace{-2.5cm}
{Fig. S6. Volume-averaged magnetic field strength at present for
four temperature phases of the IGM and for all the gas from structure
formation simulations with different resolutions of $N_{\rm cell}^3$.}
\end{figure}

\clearpage

\noindent{\bf References}

\begin{enumerate}

\item[S1.]
D. Ryu, J. P. Ostriker, H. Kang, R. Cen,
{\it Astrophys. J.} {\bf 414}, 1 (1993)

\item[S2.]
D. Ryu, H. Kang, E. Hallman, T. W. Jones,
{\it Astrophys. J.} {\bf 593}, 599 (2003)

\item[S3.]
H. Kang, D. Ryu, R. Cen, J. P. Ostriker,
{\it Astrophys. J.} {\bf 669}, 729 (2007)

\item[S4.]
R. Cen, J. P. Ostriker,
{\it Astrophys. J.} {\bf 514}, 1 (1999)

\item[S5.]
R. Dav\'e \etal~
{\it Astrophys. J.} {\bf 552}, 473 (2001)

\item[S6.]
H. Kang, D. Ryu, R. Cen, D. Song,
{\it Astrophys. J.} {\bf 620}, 21 (2005)

\item[S7.]
W. D. Hayes,
{\it J. Fluid Mech.} {\bf 2}, {595} (1957)

\item[S8.]
D. Ryu, H. Kang
{\it J. Korean Astron. Soc.} {\bf 37}, 477 (2004)

\item[S9.]
L. Biermann,
{\it Z. Naturforsch} {\bf 5a}, 65 (1950)

\item[S10.]
R. M. Kulsrud, R. Cen, J. P. Ostriker, D. Ryu,
{\it Astrophys. J.} {\bf 480}, 481 (1997)

\item[S11.]
N. Y. Gnedin, A. Ferrara, E. G. Zweibel,
{\it Astrophys. J.} {\bf 539}, 505 (2000)

\item[S12.]
M. V. Medvedev, L. O. Silva, M. Kamionkowski,
{\it Astrophys. J.} {\bf 642}, L1 (2006)

\item[S13.]
A. R. Bell,
{\it Mon. Not. Roy. Astron. Soc.} {\bf 353}, 550 (2004)

\item[S14.]
P. Goldreich, S. Sridhar,
{\it Astrophys. J.} {\bf 438}, 763 (1995)

\item[S15.]
J. Cho, E. T. Vishniac,
{\it Astrophys. J.} {\bf 538}, 217 (2000)

\item[S16.]
J. Cho, E. T. Vishniac, A.Beresnyak, A. Lazarian, D. Ryu,
{\it Astrophys. J.} submitted (2008)

\item[S17.]
J. Cho, A. Lazarian,
{\it Phys. Rev. Lett.} {\bf 88}, 245001 (2002)

\item[S18.]
N. E. L. Haugen, A. Brandenburg, A. J. Mee,
{\it Mon. Not. Roy. Astron. Soc.} {\bf 353}, 947 (2004)

\end{enumerate}


\begin{thebibliography}{}

\bibitem{ct02}
C. L. Carilli, G. B. Taylor,
{\it Annu. Rev. Astron. Astrophys.} {\bf 40}, 319 (2002)

\bibitem{gf04}
F. Govoni, L. Feretti,
{\it Int. J. Mod. Phys. D} {\bf 13}, 1549 (2004)

\bibitem{ckb01}
T. E. Clarke, P. P. Kronberg, H. B\"ohringer,
{\it Astrophys. J.} {\bf 547}, L111 (2001)

\bibitem{rkb98}
D. Ryu, H. Kang, P. L. Biermann,
{\it Astron. Astrophys.} {\bf 335}, 19 (1998)

\bibitem{xkhd06}
Y. Xu, P. P. Kronberg, S. Habib, Q. W. Dufton,
{\it Astrophys. J.} {\bf 637}, 19 (2006)

\bibitem{sfmb04} 
P. Schuecker, A. Finoguenov, F. Miniati, H. B\"ohringer, U. G. Briel,
{\it Astron. Astrophys.} {\bf 426}, 387 (2004)

\bibitem{kcor97} 
R. M. Kulsrud, R. Cen, J. P. Ostriker, D. Ryu,
{\it Astrophys. J.} {\bf 480}, 481 (1997)

\bibitem{nvk07} 
D. Nagai, A. Vikhlinin, A. V. Kravtsov,
{\it Astrophys. J.} {\bf 655}, 98 (2007)

\bibitem{mgfg04}
M. Murgia \etal,
{\it Astron. Astrophys.} {\bf 424}, 429 (2004)

\bibitem{ve05} 
C. Vogt, T. En{\ss}lin,
{\it Astron. Astrophys.} {\bf 434}, 67 (2005)

\bibitem{rkhj03}
D. Ryu, H. Kang, E. Hallman, T. W. Jones,
{\it Astrophys. J.} {\bf 593}, 599 (2003)

\bibitem{psej06}
C. Pfrommer, V. Springel, T. A. En{\ss}lin, M. Jubelgas,
{\it Mon. Not. Roy. Astron. Soc.} {\bf 367}, 113 (2006)

\bibitem{krco07}
H. Kang, D. Ryu, R. Cen, J. P. Ostriker,
{\it Astrophys. J.} {\bf 669}, 729 (2007)

\bibitem{quest88}
K. B. Quest,
{\it J. Geophys. Res.} {\bf 93}, 9649 (1988)

\bibitem{bell78} 
A. R. Bell,
{\it Mon. Not. Roy. Astron. Soc.} {\bf 182}, 147 (1978)

\bibitem{bo78}
R. D. Blandford, J. P. Ostriker,
{\it Astrophys. J.} {\bf 221}, L29 (1978)

\bibitem{biermann50} 
L. Biermann,
{\it Z. Naturforsch} {\bf 5a}, 65 (1950)

\bibitem{weibel59} 
E. S. Weibel,
{\it Phys. Rev. Lett.} {\bf 2}, 83 (1959)

\bibitem{msk06}
M. V. Medvedev, L. O. Silva, M. Kamionkowski,
{\it Astrophys. J.} {\bf 642}, L1 (2006)

\bibitem{binn74}
J. Binney,
{\it Mon. Not. Roy. Astron. Soc.} {\bf 168}, 73 (1974)

\bibitem{dw00}
G. Davies, L. M. Widrow,
{\it Astrophys. J.} {\bf 540}, 755 (2000)

\bibitem{va99}
H. J. V\"olk, A. M. Atoyan, 
{\it Astropart. Phys.} {\bf 11}, 73 (1999).

\bibitem{kdlc01}
P. P. Kronberg, Q. W. Dufton, H. Li, S. A. Colgate, 
{\it Astrophys. J.} {\bf 560}, 178 (2001)

\bibitem{ssh06} 
K. Subramanian, K., A. Shukurov, N. E. L. Haugen, 
{\it Mon. Not. Roy. Astron. Soc.} {\bf 366}, 1437 (2006).

\bibitem{rokc93}
D. Ryu, J. P. Ostriker, H. Kang, R. Cen,
{\it Astrophys. J.} {\bf 414}, 1 (1993)

\bibitem{kz07}
R. M. Kulsrud, E. G. Zweibel,
{\it Rep. Prog. Phys.} {\bf 71}, 046901 (2008)

\bibitem{cv00}
J. Cho, E. T. Vishniac,
{\it Astrophys. J.} {\bf 538}, 217 (2000)

\bibitem{gbf04}
B. M. Gaensler, R. Beck, L. Feretti,
{\it New Astron. Rev.} {\bf 48}, 1003 (2004)

\bibitem{dkrc08}
S. Das, H. Kang, D. Ryu, J. Cho
{\it Astrophys. J.} {\bf 681}, in press (2008)

\end{thebibliography}
\end{document}